\documentclass[reprint,superscriptaddress,amsmath,amssymb,aps,prl]{revtex4-1}

\usepackage{graphicx}
\usepackage{dcolumn}
\usepackage{bm}
\usepackage[colorlinks=true,urlcolor=blue,linkcolor=blue,citecolor=blue]{hyperref}
\usepackage[normalem]{ulem}
\usepackage{tikz}
\usetikzlibrary{quantikz}
\usepackage{amsmath}
\usepackage{amssymb}
\usepackage{bbold}
\DeclareMathOperator{\Tr}{Tr}
\DeclareUnicodeCharacter{2009}{\,} 
\DeclareUnicodeCharacter{03B2}{\ensuremath{\beta}}

\usepackage[all]{hypcap} 
\usepackage{braket}

\newcommand{\customref}[2]{\hyperref[#1]{\ref*{#1}#2}}

\usepackage{xcolor}

\definecolor{Ured}{HTML}{FF5C5C}
\definecolor{Ublue}{HTML}{ADD8E6}
\definecolor{Ugreen}{HTML}{198a11}

\renewcommand{\vec}[1]{\boldsymbol{#1}}

\begin{document}

\title{Accurate optimal quantum error correction thresholds from coherent information}

\author{Luis Colmenarez}
\email{colmenarez@physik.rwth-aachen.de}
\affiliation{Institute for Quantum Information, RWTH Aachen University, 52056 Aachen, Germany}
\affiliation{Institute for Theoretical Nanoelectronics (PGI-2), Forschungszentrum Jülich, 52428 Jülich, Germany}

\author{Ze-Min Huang}
\affiliation{Joint Quantum
Institute, University of Maryland, College Park, Maryland 20742, USA}
\affiliation{Institute for Theoretical Physics, University of Cologne, 50937 Cologne, Germany}

\author{Sebastian Diehl}
\affiliation{Institute for Theoretical Physics, University of Cologne, 50937 Cologne, Germany}

\author{Markus M\"{u}ller}%
\affiliation{Institute for Quantum Information, RWTH Aachen University, 52056 Aachen, Germany}
\affiliation{Institute for Theoretical Nanoelectronics (PGI-2), Forschungszentrum Jülich, 52428 Jülich, Germany}

\date{\today}

\begin{abstract}

Quantum error correcting (QEC) codes protect quantum information from decoherence, as long as error rates fall below critical error thresholds. In general, obtaining thresholds implies simulating the QEC procedure using, in general, sub-optimal decoding strategies. In a few cases and for sufficiently simple noise models, optimal decoding of QEC codes can be framed as a phase transition in disordered classical spin models. In both situations, accurate estimation of thresholds demands intensive computational resources. Here we use the coherent information of the mixed state of noisy QEC codes to accurately estimate the associated optimal QEC thresholds already from small-distance codes at moderate computational cost. We show the effectiveness and versatility of our method by applying it first to the topological surface and color code under bit-flip and depolarizing noise. We then extend the coherent information based methodology to phenomenological and quantum circuit level noise settings. For all examples considered we obtain highly accurate estimates of optimal error thresholds from small, low-distance instances of the codes, in close accordance with threshold values reported in the literature. Our findings establish the coherent information as a reliable competitive \textit{practical} tool for the calculation of optimal thresholds of state-of-the-art QEC codes under realistic noise models.

\end{abstract}
\maketitle

Successful quantum error correction (QEC) is a fundamental milestone on the route to fault-tolerant quantum computing \cite{terhal_quantum_2015,nielsen_quantum_2010,georgescu_25_2020}. QEC codes offer protection against decoherence by encoding quantum information redundantly in entangled multi-qubit states. The detrimental effect of noise can then be counteracted by measuring quantum correlations. These measurement outcomes, forming the error syndrome, can be processed by decoding procedures to correct errors and thereby restore the encoded quantum states. It is known 
that there are fundamental limits to the amount of noise that QEC codes can tolerate \cite{shor_scheme_1995}. There is a \emph{threshold} error rate beyond which the quantum information can not be retrieved with certainty \cite{aharonov_fault-tolerant_2008,knill_resilient_1998,kitaev_fault-tolerant_2003,shor_fault-tolerant_1996}. To be concrete, below threshold the originally prepared state can be recovered with fidelity approaching one \cite{schumacher_quantum_1996,knill_theory_1997}. Therefore computing and/or estimating QEC thresholds is essential for developing and improving QEC codes. Recent QEC experiments, e.g.~with trapped ions \cite{nigg_quantum_2014,postler_demonstration_2022,ryan-anderson_implementing_2022}, superconducting qubits \cite{krinner_realizing_2022,zhao_realization_2022,andersen_repeated_2020,kelly_state_2015,google_quantum_ai_exponential_2021,google_quantum_ai_suppressing_2023} and neutral atoms \cite{bluvstein_logical_2023} have demonstrated concepts of fault-tolerant QEC, with the performance of logical qubits approaching or reaching break-even with physical qubits.

In general, the determination of error thresholds is a computationally hard problem \cite{iyer_hardness_2015}. The standard approach involves simulating the QEC procedure by sampling errors from the error model and estimating logical failure rates. The latter requires to choose a decoder \cite{lidar_quantum_2013} that detects and corrects errors. For stabilizer QEC codes and Clifford noise models, one can simulate very large code instances, allowing for good extrapolation of the threshold. Regarding the choice of decoder, maximum likelihood decoding (MLD) is optimal and thus achieves optimal thresholds~\cite{iyer_hardness_2015,fuentes_degeneracy_2021}. Therefore, thresholds from QEC simulations are optimal if and only if the decoding strategy is equivalent to MLD. One downside of MLD, in general, is its exponential scaling in the number of physical qubits for finding the optimal recovery operation given a syndrome measurement \cite{iyer_hardness_2015}, which often severely limits its practicality. In contrast, sup-optimal decoders may be efficient but do not reach optimal thresholds because they can not make use the whole information provided by the syndrome measurements.

For certain QEC codes and error models, the optimal decoding problem acquires enough structure to allow for a mapping to partition functions of statistical mechanics problems. In seminal work~\cite{dennis_topological_2002} it was shown that the threshold problem of the two-dimensional (2D) toric code \cite{kitaev_quantum_1997} is equivalent to the phase transition of the classical Random Bond Ising model (RBIM) in 2D along the Nishimori line \cite{nishimori_internal_1981}. Since then, statistical mechanics mappings for other topological codes and error models have been developed \cite{chubb_statistical_2021,wang_confinement-higgs_2003,kubica_three-dimensional_2018,bombin_strong_2012,ohzeki_accuracy_2009,katzgraber_error_2009,katzgraber_stability_2013,vodola_fundamental_2022,fan_diagnostics_2023,wang_intrinsic_2023}. This approach has the advantage of being tractable via Monte Carlo sampling of the partition function, providing estimates of QEC thresholds as critical points of phase transitions. The main challenge lies in the significant amount of computational resources needed to accurately locate critical points of disordered interacting spin models. In summary there are two ways to estimate optimal thresholds of QEC codes: i) a general QEC simulation and use of MLD \cite{iyer_hardness_2015,fuentes_degeneracy_2021} and ii) constructing statistical mechanics mappings for specific codes and error models \cite{dennis_topological_2002,chubb_statistical_2021,wang_confinement-higgs_2003,kubica_three-dimensional_2018,bombin_strong_2012,ohzeki_accuracy_2009,katzgraber_error_2009,katzgraber_stability_2013,vodola_fundamental_2022,fan_diagnostics_2023,lee_quantum_2023,wang_intrinsic_2023}.

In this work we introduce a third approach to the estimation of optimal QEC thresholds based on the \emph{coherent information} (CI) \cite{schumacher_quantum_1996,lloyd_capacity_1997}. This quantity is the theoretical upper limit to the amount of quantum information that can be recovered from a noisy mixed state, hence reflecting the capability of the given QEC code to protect the logical qubits under a particular error model. As such, the CI signals the threshold error rate beyond which the quantum information can not be recovered by any means. Thus only optimal decoders can attain the threshold obtained from the CI, while thresholds reached by sub-optimal decoders are always smaller than the optimal thresholds we study in this work.
To illustrate the use and power of the CI, we estimate optimal thresholds of topological QEC codes in the \emph{code capacity} setting, namely, when the state becomes corrupted after a single round of noise on each data qubit. We show that the CI, evaluated for small instances of topological codes, provides threshold values that come very close to known optimal thresholds, exhibiting strikingly small finite size effects. Moreover, in real devices all components of the circuit are faulty (e.g. measurement and gates), which compels the inclusion of gate errors and stabilizer measurements in the error model. Therefore, we adapt the CI as a tool to determine critical error thresholds to the \emph{phenomenological and circuit level noise} settings. Formally, the latter setting includes gate and measurement errors alike, whereas the former only considers measurement errors. We then apply this approach to the quantum repetition code in the aforementioned settings and show that this again yields very accurate threshold estimates. Importantly, in all cases of study the finite distance crossings are \emph{almost exactly} at the location of the known thresholds (discrepancies start at the third digit), suggesting little finite size corrections in the CI.

Before introducing formal definitions, we briefly compare the approach introduced here with the other two main approaches for obtaining optimal thresholds. On the one hand, statistical mechanics mappings must be crafted on a case-by-case basis and are limited mostly to the code capacity and phenomenological noise settings - with few exceptions  ~\cite{vodola_fundamental_2022,chubb_statistical_2021}. In contrast, the CI approach is easily adapted to different codes and error models. On the other hand, MLD and computing the CI both suffer from exponential overhead in the number of physical qubits. However the CI does not require error sampling, thus getting rid of statistical errors in the threshold estimation, and is thus purely limited by minimal finite size-effects.

\begin{figure}
    \centering
    \includegraphics[width=0.45\textwidth]{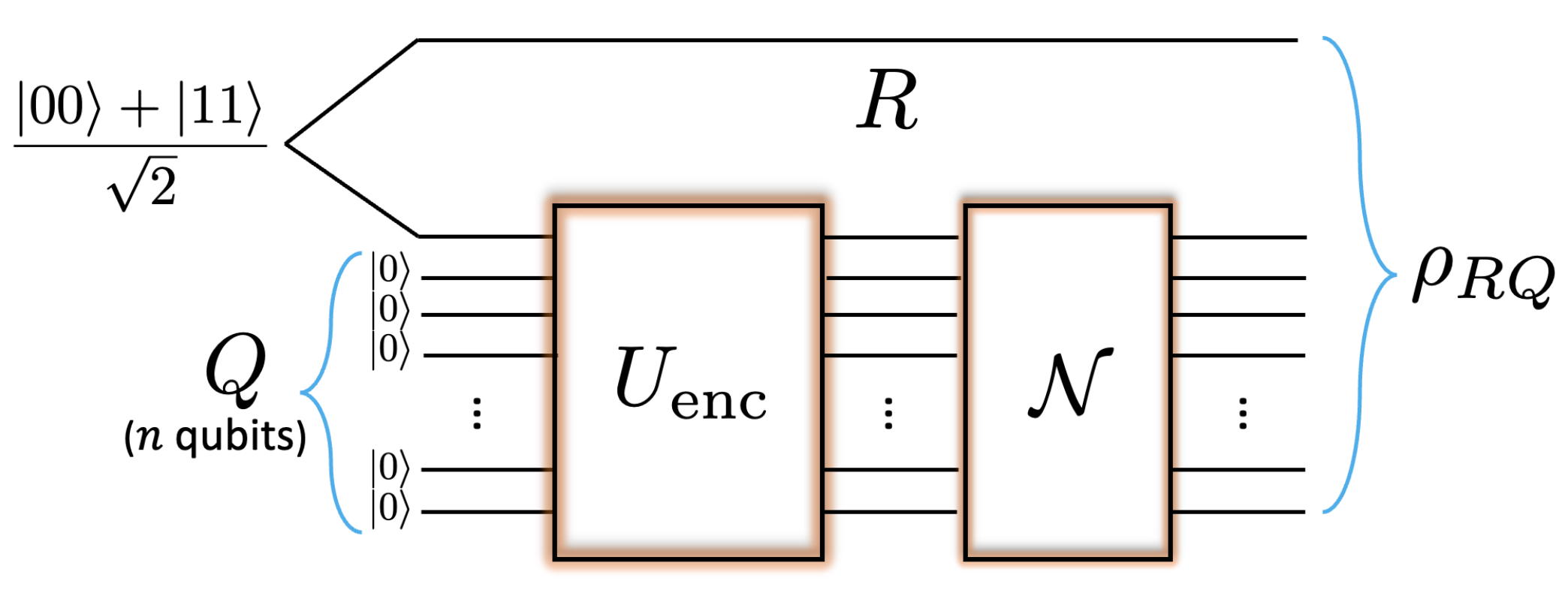}
    \caption{Setup for the coherent information for a QEC code encoding a single logical qubit. We start with a Bell pair between two physical qubits, one of those qubits is kept as a (noise-free) reference qubit ($R$). The other qubit is used for preparing the logical state on the register of $n$ qubits ($Q$). The encoding unitary $U_{\text{enc}}$ on the data qubits creates a superposition of logical states, such that in the $RQ$ system we have $(|0_R0_L\rangle+|1_R1_L\rangle)/\sqrt{2}$. After encoding, the $Q$ system is exposed to noise $\mathcal{N}$ according to a noise model. The coherent information of the noisy state $\rho_{RQ}$ allows one to extract optimal thresholds. }
    \label{fig:coherent_information_set_up}
\end{figure}

\emph{The coherent information (CI)} was first introduced as a measure of the quantum information transmitted through a noisy channel \cite{lloyd_capacity_1997}. Indeed maximizing the CI over all possible encodings yields the quantum capacity of a noisy quantum channel~\cite{gyongyosi_survey_2018}. At the same time it was established as key ingredient for the existence of a successful quantum error correction protocol \cite{schumacher_quantum_1996,barnum_information_1998}. Following the setting shown in Fig.~\ref{fig:coherent_information_set_up}, the coherent information is defined as 
\begin{equation}\label{eq:def_coherent_information}
    I = S(\rho_Q) - S(\rho_{RQ}).
\end{equation}
Here, $S(\rho)=-\Tr{(\rho\log\rho)}$ is the von Neumann entropy of the state and $\rho_Q = \Tr_R{(\rho_{RQ})}$ is the reduced density matrix after tracing out the reference qubit. It was shown in Ref.~\cite{schumacher_quantum_1996} that $I$ can only decrease or stay constant after successively applying CPTP maps to the $Q$ system. In this setup the CI before application of the noise is $I_0=\log 2$, therefore $I\leq \log2$. The equality $I=\log2$ ensures the existence of a quantum error correction procedure that recovers the state with unit fidelity. In Ref.~\cite{divincenzo_quantum-channel_1998}, the CI has been used in finding thresholds of concatenated repetition codes. Although there the coherent information was identified as the quantum code capacity, the QEC threshold is then defined as the error rate for such quantity to vanish for any code distance. Generalizing the setup in Fig.~\ref{fig:coherent_information_set_up} to codes with $k>1$ logical qubits is straightforward, see Supplemental Material (SM) for details.

\emph{Error models}: We focus on two types of single qubit error models: bit-flip noise and depolarizing noise. The bit-flip channel on qubit $i$ with error rate $p$ reads $\mathcal{N}^{b}_i(\rho) = (1-p)\rho + pX_i\rho X_i $, while the depolarizing channel is written as $\mathcal{N}^{d}_i(\rho) = (1-p)\rho + \dfrac{p}{3}X_i\rho X_i+\dfrac{p}{3}Y_i\rho Y_i+\dfrac{p}{3}Z_i\rho Z_i $. All qubits are exposed to the same error channel such that $\mathcal{N} = \prod_{i=1}^{n}\mathcal{N}^{a}_i$, where $a=b,d$ denotes the type of channel.

\emph{Quantum codes and their thresholds}: We choose to benchmark our method for two well-known topological stabilizer codes, (i) the rotated surface code and (ii) the topological color code, both being leading contenders for practical QEC. 
(i) The rotated surface code 
$[[n,1,\sqrt{n}]]$ \cite{bombin_optimal_2007,tomita_low-distance_2014}, which is asymptotically equivalent to the toric code \cite{kitaev_quantum_1997}, encodes $k=1$ logical qubit in $n$ physical qubits and has a code distance (minimal support of logical operators) of $d=\sqrt{n}$. In Ref.~\cite{dennis_topological_2002} the threshold of the toric code under bit-flip noise was estimated from mapping the QEC code to the 2D RBIM as already mentioned, yielding a threshold $p_{th}=0.109(2)$. Depolarizing noise, which also contains the occurrence of $Y$-type errors, which trigger both non-trival $X$- and $Z$-type syndromes, introduces correlations between bit-flip and dephasing noise. Thus the corresponding statistical mechanics model becomes a stack of two RBIMs on square lattices with inter-layer 4-body interactions \cite{bombin_strong_2012}. The numerically estimated threshold under depolarizing noise is $p_{th}=0.189(3)$ \cite{bombin_strong_2012}. (ii) The triangular 4.8.8  color code $[[n(d),1,d]]$ encodes $k=1$ logical qubit in $n(d)=(d^2-1)/2+d$ physical qubits \cite{parrado-rodriguez_rescaling_2022} and belongs to the family of 2D topological color codes \cite{bombin_topological_2007}. The reported threshold under bit-flip noise only is $p_{th}\approx0.109$ \cite{ohzeki_accuracy_2009,katzgraber_error_2009,ohzeki_locations_2009,de_queiroz_location_2009,hasenbusch_multicritical_2008}. Thus it is widely believed that the toric and 2D color codes have the same code capacity threshold under bit flip noise. As for the surface code, depolarizing noise introduces error correlations that manifest themselves as inter-layer interaction between two originally independent 2D disordered spin models. The threshold for the 4.8.8 color codes under depolarizing noise was estimated as $p_{th}=0.189(2)$ \cite{bombin_strong_2012}. Details about the rotated surface and 4.8.8 color code are given in the SM.

\begin{figure}[h]
    \centering
    \includegraphics{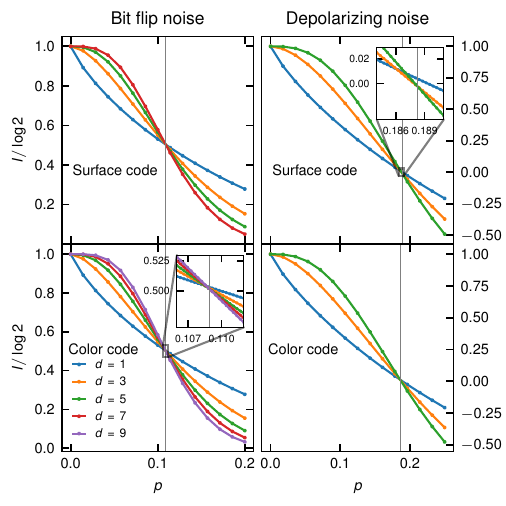}
    \caption{Coherent information for surface (upper row), color code (lower row) under bit flip noise (left column) and depolarizing noise (right column). We show distances $d=3$ to $9(7)$ for 4.8.8 color (rotated surface) code under bit-flip noise and $d=3,5$ for both codes under depolarizing noise. $d=1$ denotes the CI of a single qubit. Grey vertical lines are the crossing between the two largest distances available (see Table~\ref{tab:thresholds}).}
    \label{fig:topo_codes}
\end{figure}

\begin{figure}[h]
    \centering
    \includegraphics{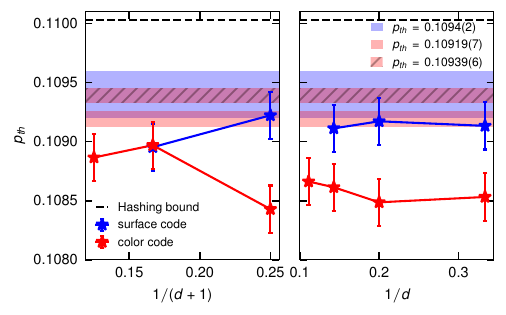}
    \caption{Left: Finite distance crossings of CI between consecutive distances $d\geq 3$ and $d+2$ for color and surface code under bit-flip noise. Right: Crossing between surface and color codes with distance $d\geq3$ and the single qubit CI. Shaded regions are previous estimates of thresholds: $p_{th}=0.1094(2)$ from Ref. \cite{dennis_topological_2002} for the surface code. For the color code we show $p_{th}=0.10919(7)$ from Ref. \cite{hasenbusch_multicritical_2008} and $p_{th}=0.10939(6)$ from Ref. \cite{de_queiroz_location_2009}. Error bars are the error rate resolution prior to fitting \cite{crossings}.}
    \label{fig:topo_codes_crossings}
\end{figure}

\begin{table}
\begin{center}
\begin{tabular}{ |c|c|c|c|c| } 
 \hline
  & Bit-flip (CI) & Depol. (CI) & Bit flip $p_{th}$ & Depol. $p_{th}$  \\ 
 \hline
 Surface & 0.1089(2) & 0.1882(2) & 0.1094(2) \cite{dennis_topological_2002} & 0.189(3) \cite{bombin_strong_2012} \\ 
 \hline
 Color & 0.1088(2) & 0.1862(2) & 0.10919(7) \cite{hasenbusch_multicritical_2008} & 0.189(2) \cite{bombin_strong_2012}  \\ 
 \hline
\end{tabular}
\end{center}
\caption{Crossings of coherent information (CI) between the two largest distance available \cite{crossings} plotted in Fig.~\ref{fig:topo_codes}. For comparison, we show the known thresholds for surface and 2D color codes.}
\label{tab:thresholds}
\end{table}

\emph{Coherent information in topological codes}: To reach codes with relatively large number of data qubits we exploit the structure of the mixed state and properties of the CI and error model. This allows us to numerically compute the CI for codes with up to $n=49$ data qubits ($d=9$ color code) under bit-flip noise and $n=25$ for depolarizing noise ($d=5$ surface code). For details about the numerical calculations see the SM. It is important to note that we exploit general properties of stabilizer states and Pauli noise, thus we expect the benefits from such properties to hold regardless of other independent factors, e.g. spatial dimensionality and error correlations. In Fig.~\ref{fig:topo_codes} we show the CI for surface and color code alongside the CI of a single qubit ($d=1$).  The CI of the code with distance $d\geq5$ crosses the curve for distance $d-2$ almost at the same error rate. Indeed the crossing point of the two largest available distances is equal to the known values of the threshold up to two decimal digits (see Table~\ref{tab:thresholds}). Given the smallness of the code instances, it is quite remarkable that the finite size crossings yield such accurate estimates of the QEC threshold. Furthermore we present a pseudo-threshold analysis by plotting the crossings of the CI of codes with distance $d>1$ with the single qubit CI (see Fig.~\ref{fig:topo_codes_crossings}). We see a slight non-monotonous trend of the pseudo-thresholds which may be caused by finite size effects. However, it is worth to note that the finite size effects are noticeable only from the third decimal digit onwards (see insets of Fig.~\ref{fig:topo_codes}).

\begin{figure*}
  \includegraphics[width=0.85\textwidth]{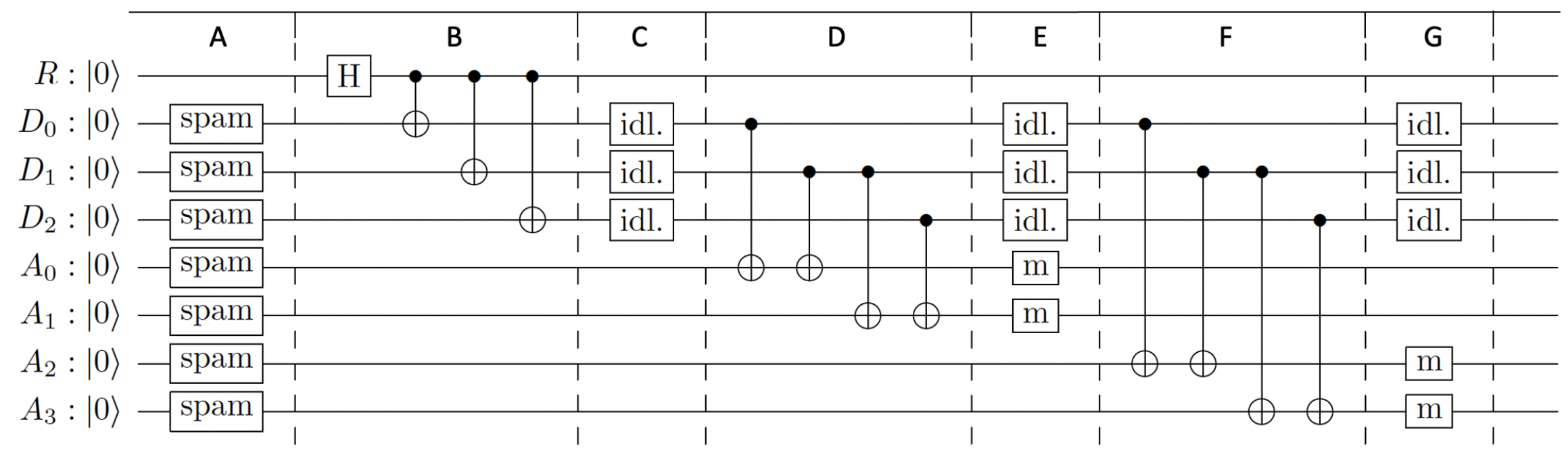}
  \caption{Setup for computing coherent information with circuit level noise in the three-qubit repetition code. The qubits $D_i$ and $R$ are the data and reference qubits respectively. The qubits $A_i$ are ancilla qubits that store the syndrome measurement. Part \textbf{A} denotes state preparation and measurement (SPAM) errors with probability $p_{sp}$. In \textbf{B} the state $(|0_R0_L\rangle+|1_R1_L\rangle)/\sqrt{2}$ is prepared. This part is assumed to be noiseless. Parts \textbf{C},\textbf{E} and \textbf{G} contain idling noise with probability $p_{id}$. In parts \textbf{E} and \textbf{G} measurement errors are modeled as bit-flip channel with probability $p_m$. In parts \textbf{D} and \textbf{F} the stabilizer information is coherently mapped to the auxiliary qubits (without actually projectively measuring the auxiliary qubits). The CNOT gates in the stabilizer measurements fail with probability $p_2$ (see SM for details about the error model). Then the data qubits are exposed to bit-flip noise as usual. Here we call phenomenological noise the case $p_{sp}=p_2=0$ and $p_m=p_{id}$, while the setting $p_{sp}=p_m=p_{id}=p_{2}=\lambda$ represents a circuit level noise realization.}
    \label{fig:circuit_repetition_code}
\end{figure*}

\begin{figure}
    \centering
    \includegraphics[width=0.5\textwidth]{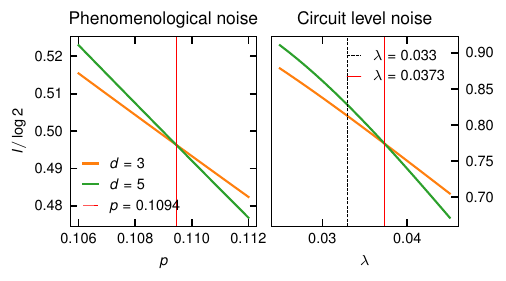}
    \caption{Coherent information for the data and auxiliary qubit state in the repetition code for $d=3,5$. Left: result for phenomenological noise; the red line denotes the crossing at $p = 0.1094(2)$. Right: result for circuit level noise; the red line denotes the crossing at $\lambda = 0.0373(5)$; the blacked dashed line is the threshold reported in \cite{rispler_towards_2020} using MLD.}
    \label{fig:repetition_code}
\end{figure}

\emph{Phenomenological and circuit level noise}: So far we have considered the code capacity setting only. Now we examine the more realistic case when measurements and gates are also faulty. To this end, we consider the $n$-qubit repetition code in the $Z$ basis exposed to bit-flip noise. This code has distance $d=n$ but protects only against bit-flip noise. Faulty measurements of syndrome qubits (phenomenological noise) are modeled by bit-flip channels on the ancilla qubits right after each round of stabilizer measurements (see setting and error model in Fig.~\ref{fig:circuit_repetition_code}). In this setting the repetition code asymptotically behaves as a surface code in a space-time description \cite{dennis_topological_2002}. When data qubit and measurement errors occur with the same error rate, the threshold is the same found for the surface code under bit-flip noise $p_{th}\approx0.109$ (see Table~\ref{tab:thresholds}). The CI crossing between $d=3$ and $d=5$ repetition code under phenomenological noise is located at $p\approx0.1094(2)$ (see Fig.~\ref{fig:repetition_code}) which is \emph{precisely} the aforementioned value \cite{dennis_topological_2002}.  We also study the more realistic case of circuit level noise, namely considering gate errors in the syndrome measurement and errors in the initialization of qubits (SPAM errors). The extraction of fundamental thresholds for quantum circuit noise settings is challenging to tackle by mappings to disordered classical spin models \cite{chubb_statistical_2021,venn_coherent-error_2023}, but was considered previously in ~\cite{rispler_towards_2020} for a quantum repetition code. When all circuit error rates (SPAM, idling, gate and measurement errors) are equal to $\lambda$, the threshold was computed using MLD, and found to lie at $\lambda_{th}\approx 0.033$. Later, statistical mechanics mappings for circuit level noise in repetition codes were developed \cite{vodola_fundamental_2022}, confirming the previously found threshold. Our CI-based analysis shows a crossing at $\lambda \approx 0.0373(5)$. The difference starts at the third digit, compatible within error bars with the previous studies. 

Now we explain how to generalize the CI to incorporate noisy syndrome measurements. We start off by preparing the logical states and expose the data qubits to noise. Then the syndrome is measured $d-1$ times via ancilla couplings (see the illustration of the $n=3$ setup and explanation of the error model in Fig.~\ref{fig:circuit_repetition_code}). After the last step, we consider the whole quantum state -- consisting of the reference qubit, the data and the ancilla qubits -- as the state $\rho_{RQ}$ and $\rho_Q$ as the reduced state of data and ancilla qubits only, after tracing out the reference qubit. The coherent information is defined as usual, $I = S(\rho_Q)-S(\rho_{RQ})$. The total number of qubits involved is $1+d+(d-1)^2$, thus our numerical calculation is limited to $d\leq5$. Unlike previous works \cite{rispler_towards_2020,vodola_fundamental_2022}, we model all sources of errors as bit-flip channels, instead of depolarizing channel. Since phase errors inherently destroy entanglement in the repetition code and the CI tests the ability to reconstruct entanglement, an extensive number of depolarizing locations in the circuit inevitably shifts the threshold towards zero. In Refs.~\cite{rispler_towards_2020,vodola_fundamental_2022} the authors recover a finite threshold under depolarizing noise because they test the ability to reconstruct one code word, e.g. $|0\rangle_L$, instead of rebuilding the state $(|0_R0_L\rangle+|1_R1_L\rangle)/\sqrt{2}$. Despite their different underlining principle, both procedures are equivalent in terms of QEC capabilities of quantum codes \cite{lloyd_capacity_1997,schumacher_quantum_1996,barnum_information_1998}. Besides, in order to make a fair comparison with previous works 
\cite{rispler_towards_2020,vodola_fundamental_2022}, we do not include errors in the encoding circuit of the respective logical states.

\emph{Discussion}: In this work we have established the CI in small codes as a competitive \textit{practical} tool for reliably estimating optimal QEC thresholds with high precision on the basis of a number of different error models. The efficiency of the approach is due to our observation of very tiny finite size effects, which in light of the variety of codes investigated here seems general, and at any rate is testable in practice.

It is worth to discuss the computational cost of CI and Monte Carlo based methods. The CI is a well-defined non-linear function of the whole mixed-state density matrices, therefore it can not be computed using trajectory approaches, e.g. Clifford simulations, and requires in general to allocate $\mathcal{O}(e^{n})$ parameters; limiting its computation to few tens of qubits. However, we have shown that the exponential scaling can be eased by exploiting the structure of the state and error model, allowing us to explore code instances beyond the smallest distances. In the present work, the CI on the largest relevant Hilbert space, surface code $[[25,1,5]]$, was computed on a laptop over one hour (for a fixed error rate). On the contrary, determining the phase diagram of disordered spin models corresponding to noisy QEC codes, for a fixed error rate, e.g.~using tempering Monte Carlo \cite{bombin_strong_2012,kubica_three-dimensional_2018}, requires intensive computational resources and longer times to achieve thresholds with similar level of accuracy compared to the use of the CI (see SM for computing time of the CI). Besides, the CI is free of stochastic errors, unlike MLD that needs sampling from an error model. The same analysis applies when considering phenomenological and circuit level noise because the ancilla qubits cause the same effect on the three methods. To be precise, they add up to the exponential scaling of MLD and CI and add one extra dimension to the statistical mechanics mappings \cite{dennis_topological_2002,wang_confinement-higgs_2003}.

We thus envision the CI as a promising tool for estimating optimal thresholds in regimes that are today difficult to access. One case is circuit level noise where statistical mechanics mappings are rare  \cite{vodola_fundamental_2022}, including coherent noise \cite{bravyi_correcting_2018,chubb_statistical_2021,venn_coherent-error_2023}. Another potential application case are new codes, most prominently quantum low-density parity check (qLDPC) codes \cite{breuckmann_quantum_2021} (see the SM for a calculation of CI in a specific qLDPC code), whose growing number of logical qubits limits the direct use of statistical mechanics mappings \cite{placke_random-bond_2023} and MLD \cite{cao_qecgpt_2023}. Hence, the CI provides an unbiased method to estimate QEC thresholds for codes with multiple logical qubits and for general noise models, ranging from code capacity and phenomenological noise to circuit level noise.

\begin{acknowledgments}
We thank David DiVincenzo, Seyong Kim and Josias Old for useful discussions. L.C. and M.M. gratefully acknowledge funding by the U.S. ARO Grant No. W911NF-21-1-0007. M.M. furthermore acknowledges funding from the European Union’s Horizon Europe research and innovation programme under grant agreement No 101114305 (“MILLENION-SGA1” EU Project), and this research is also part of the Munich Quantum Valley (K-8), which is supported by the Bavarian state government with funds from the Hightech Agenda Bayern Plus. M.M. acknowledges funding from
the ERC Starting Grant QNets through Grant No. 804247, and from the European Union’s Horizon Europe research and innovation program under Grant Agreement No. 101046968 (BRISQ). 
M.M. also acknowledges support for the research that was sponsored by IARPA and the Army Research Office, under the Entangled Logical Qubits program, and was accomplished under Cooperative Agreement Number W911NF-23-2-0216. The views and conclusions contained in this document are those of the authors and should not be interpreted as representing the official policies, either expressed or implied, of IARPA, the Army Research Office, or the U.S. Government. The U.S. Government is authorized to reproduce and distribute reprints for Government purposes notwithstanding any copyright notation herein. Z.-M.H. acknowledges the support from the JQI postdoctoral fellowship at the University of Maryland. S.D. and M.M. acknowledge support from the Deutsche Forschungsgemeinschaft (DFG, German Research Foundation) under  Germany’s Excellence Strategy Cluster of Excellence Matter and Light for  Quantum Computing (ML4Q) EXC 2004/1 390534769, and S.D. furthermore by the DFG Collaborative  Research Center (CRC) 183 Project No. 277101999 - project B02. 
\end{acknowledgments}

\bibliography{coherent_information,footnotes}

\begin{thebibliography}{62}
\bibitem [{cro()}]{crossings}%
  \BibitemOpen
  \href@noop {} {}\bibinfo {note} {Crossings are computed by finding the root
  of the interpolated function $y=I_{d+2}-I_{d}$. The sub-index refers to the
  code distance. The root finder routine used was scipy.optimize.brent, which
  delivers close-to machine precision roots. Error bars are estimated by the
  resolution in error rate $\Delta p\sim 0.0002$ prior to interpolation. It
  should be noted that this is likely overestimating the interpolation error in
  finding the crossing}\BibitemShut {NoStop}%
\end{thebibliography}

\clearpage

\section{Supplemental Material}

\subsection{Topological codes considered in this work}

\begin{figure}[h]
    \centering
    \includegraphics[width=0.46\textwidth]{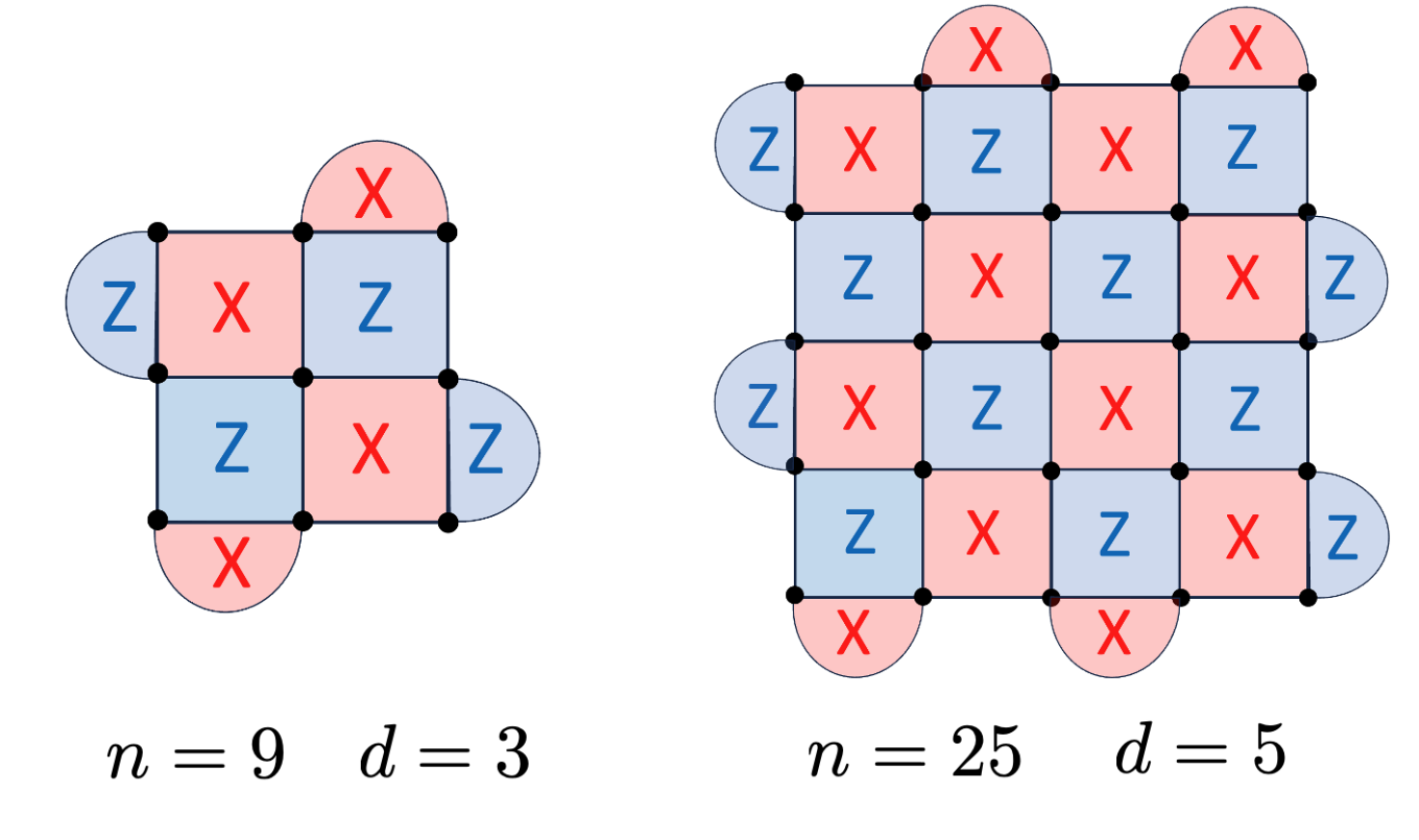}
    \caption{Small instances of rotated surface code: $d=3$ and $d=5$. In general the code distance is $d=\sqrt{n}$, and consequently $[[n=d^2,1,d]]$. Blue (Red) plaquettes correspond $Z$ ($X$) stabilizers. Ovals denote the weight-2 stabilizers while the square tiles are the weight-4 stabilizers in the bulk. Physical qubits (black dots) are located on the vertices of the square lattice.}
    \label{fig:rotated_surface_codes}
\end{figure}

\begin{figure}[h]
    \centering
    \includegraphics[width=0.485\textwidth]{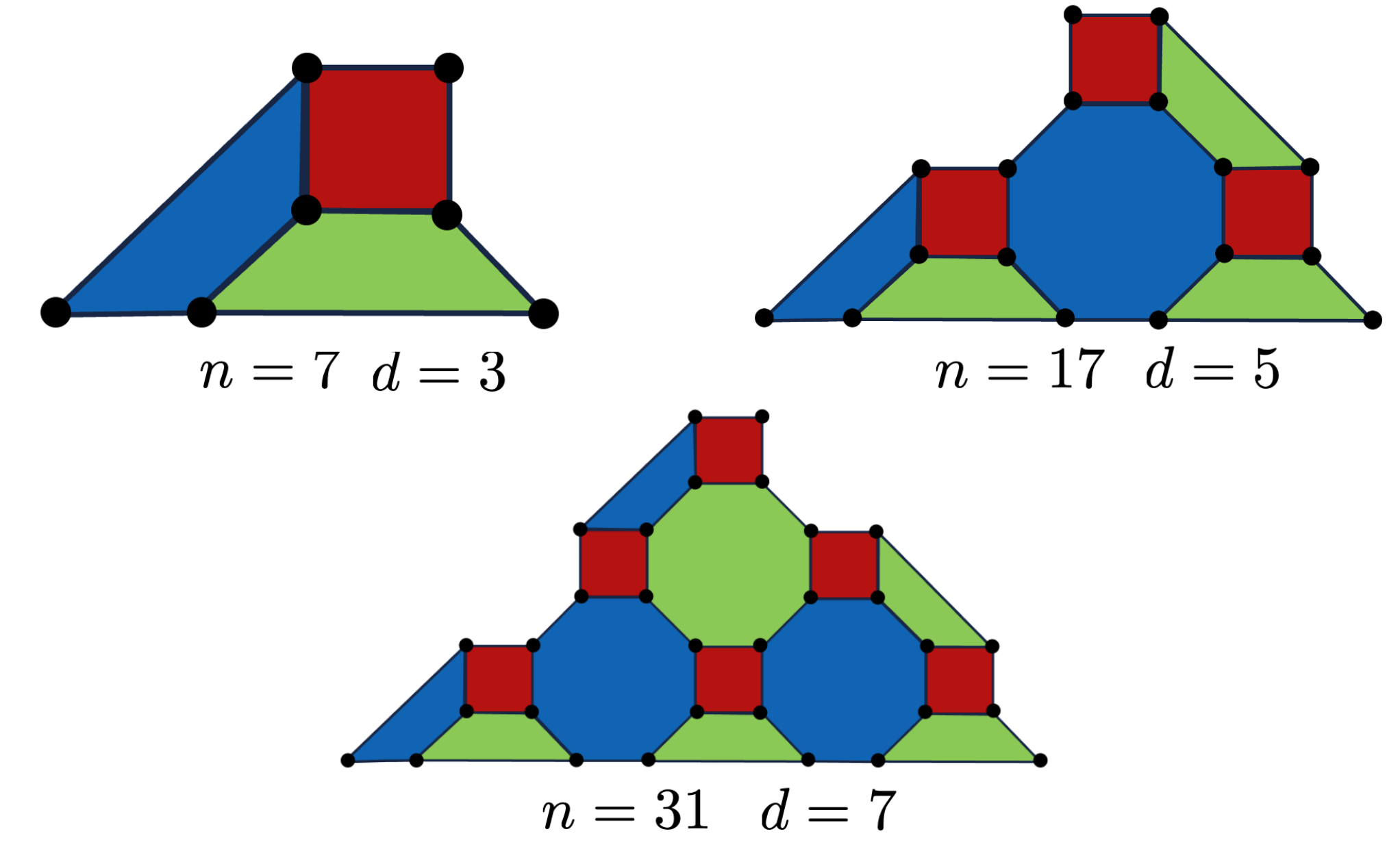}
    \caption{Low-distance code instances of triangular 4.8.8 color code: distances $d=3,5,7$. In general $n=(d^2-1)/2+d$ yielding $[[(d^2-1)/2+d,1,d]]$. Both $X$ and $Z$ stabilizers are defined on the colored plaquettes of the lattice. Physical qubits (black dots) are placed on the vertices such that each qubit is shared by at most three plaqettes. }
    \label{fig:octogonal_color_codes}
\end{figure}

In this section we are going to describe in more detail the two classes of topological codes studied in the main text. Topological codes are a broad family of codes that store quantum information in degenerate topologically ordered states \cite{bombin_introduction_2013}, thus they are the prototypical example of stabilizer code families. Surface \cite{kitaev_quantum_1997,kitaev_fault-tolerant_2003,dennis_topological_2002} an 2D color codes \cite{bombin_topological_2006,bombin_topological_2007} have received a lot of attention in recent years due to their potential in realizing fault-tolerant quantum computing \cite{postler_demonstration_2022,google_quantum_ai_suppressing_2023}. To test the coherent information, we choose representatives of surface and 2D color codes with the more gentle scaling of physical qubits and code distance. 

The rotated surface code \cite{bombin_optimal_2007,tomita_low-distance_2014} is defined on a square lattice of length $L$. The code distance $d=L$ is equal to the length of the boundary (see Fig.~\ref{fig:rotated_surface_codes}). The weight-4 stabilizers are placed on the lattice's plaquettes ($X$ and $Z$ alternating) while weight-2 stabilizers are defined on the boundaries of the lattice. The location of weight-2 operators defines two classes of boundaries: the borders with $X$ ($Z$) stabilizers is where the $Z_L$ ($X_L$) logical operator is defined. 

The $4.8.8$ color code \cite{parrado-rodriguez_rescaling_2022} is defined on a three-colorable lattice filled by a 4.8.8 tiling (see small patches of the lattice/code in Fig.~\ref{fig:octogonal_color_codes}). The number of physical qubits scales as $n=(d^2-1)/2+d$. The $d=3$ code is equivalent to the 7-qubit Steane code \cite{steane_multiple-particle_1997, postler_demonstration_2022}. Each boundary only contains plaquettes of two colors, for instance in Fig.~\ref{fig:octogonal_color_codes} the boundary at the bottom only has blue and green plaquettes. The lowest-weight logical operators are defined on the boundaries (see Fig. 2 in Ref.\cite{gutierrez_transversality_2019}). 

\subsection{Encoding more logical qubits $k>1$ and qLDPC codes}\label{sec:qldpc}

\begin{figure}[h]
    \centering
    \includegraphics[width=0.45\textwidth]{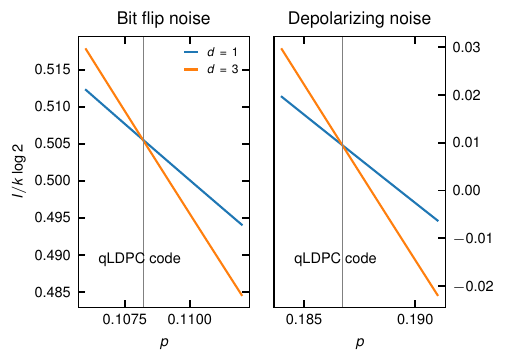}
    \caption{Coherent information for lift-twisted surface code under bit-flip (left) and depolarizing noise (right). The black dashed line is the hashing bound while the grey line is the crossing between code the code with $d=3$ and the single qubit CI. }
    \label{fig:ldpc_codes}
\end{figure}

In this section we generalize the coherent information to any number of logical qubits $k>1$ and show the CI for a qLDPC code that has $k=3$ logical qubits. Generalizing the setup in Fig.~\ref{fig:coherent_information_set_up} to codes with $k>1$ logical qubits is straightforward. One needs a generalized Bell state \cite{fujii_generalized_2001} of the $2^k$ states in the code space, for instance the state $\sum_{i}^{2^k} |i_R\rangle \otimes |i_L\rangle/2^{k/2}$ where the states $|i_R\rangle$ form an orthonormal basis on the $2^k$ Hilbert space and $|i_L\rangle$ are the same states in the code space of the given code. The CI of the noiseless state is then $I_0=k\log 2$, therefore we always refer to the normalized CI, $I/(k\log2)$. For instance for $k=2$ a standard choice of the generalized Bell state would be $|00\rangle_R|00\rangle_L+|01\rangle_R|01\rangle_L+|10\rangle_R|10\rangle_L+|11\rangle_R|11\rangle_L$ up to normalization (the tensor product symbol is omitted for simplicity). 

Quantum low-density parity-check codes (qLDPC) \cite{breuckmann_quantum_2021} is a term for all codes whose parity-checks are not extensive, namely the support of the stabilizer operators does not scale up with the size of the code, and each qubit is only involved in a constant number of stabilizers. Topological codes are a small subset of qLDPC codes where the numbers of logical qubits $k$ is fixed for any number of physical qubits $n$ in the code family. Therefore the code rate, defined as $R=k/n$, vanishes as $n\rightarrow\infty$. In order to make a more efficient use of resources, we want the ratio $k/n$ to be finite. 

As an example we choose the $[[15,3,3]]$ lift-twisted surface code \cite{old_lift-connected_2024}. In Fig.~\ref{fig:ldpc_codes} we present its CI for bit-flip and depolarizing noise. We compare the crossing of the $d=3$ code with the single qubit coherent information. The crossings lay at $p=0.1081(2)$ and $p=0.1867(2)$ for bit-flip and depolarizing noise, respectively (error bars are error rate resolution before fitting \cite{crossings}). Unfortunately, higher code distances require many more physical qubits than the ones we can currently numerically work with.

\subsection{Error model for circuit level noise}\label{sec:error_model_circuit_level_noise}

In Fig.~\ref{fig:circuit_repetition_code} we show the setup for studying circuit level and phenomenological noise. Now we describe in detail the error model used. SPAM, measurement and idling noise is modeled by single-qubit bit-flip channels, as shown in the main text, with probability $p_{sp}$, $p_{m}$ and $p_{id}$, respectively. Errors in the CNOTs used in the stabilizer measurements are modeled as a two-qubit bit-flip channel:
\begin{equation}\label{eq:two_bit_flip_channel}
\mathcal{N}_{i,j}(\rho) = (1-p_2)\rho + \dfrac{p_2}{3} X_i\rho X_i + \dfrac{p_2}{3} X_j\rho X_j + \dfrac{p_2}{3} X_i X_j\rho X_i X_j,
\end{equation}
where $i$ and $j$ are qubit indices. Let us note that we do not consider depolarizing noise as in previous works (see discussion in the main text).

\subsection{Method for computing the coherent information}

\begin{figure}[h]
    \centering
    \includegraphics[width=0.45\textwidth]{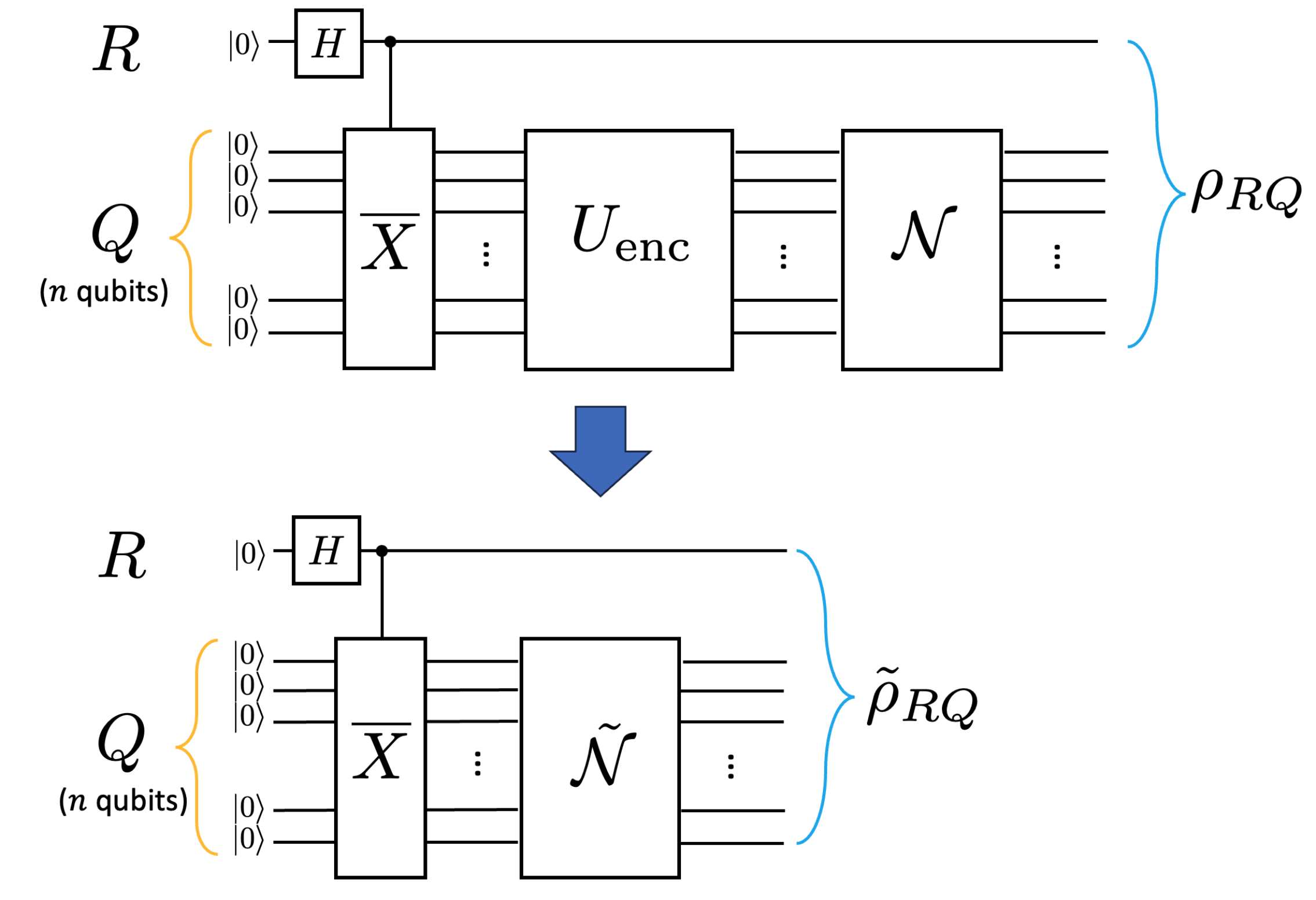}
    \caption{Schematic of transforming Pauli errors to the stabilizer basis. Each error $E$ in the error model is mapped to the stabilizer basis by propagating the error back through the encoding circuit $E U_{enc} = U_{enc} \tilde{E} $, ending up in the error model $\tilde{\mathcal{N}}$. The operator $\tilde{N}$ is introduced for ensuring $|0_L\rangle = U_{enc}|0\rangle^{\bigotimes n} $ and $|1_L\rangle = U_{enc}\overline{X}|0\rangle^{\bigotimes n} $. }
    \label{fig:error_prop}
\end{figure}

In this section we show in detail how we compute the coherent information numerically. First, we recall that we need to prepare the state
\begin{equation}\label{eq:initial_state}
|\Psi\rangle = \dfrac{|0_L0_R\rangle+|1_L1_R\rangle}{\sqrt{2}},
\end{equation}
where $|0_L\rangle$ and $|1_L\rangle$ are the logical states of the given code and $|0_R\rangle$ and $|1_R\rangle$ denote the two states of the reference qubit. This state has a simple representation in the \emph{stabilizer basis}. Let us first introduce the stabilizer basis:  
\begin{equation}\label{eq:basis}
\{|\vec{s},\vec{l},\vec{r}\rangle\}.
\end{equation}
Here $\vec{s}=(s_1,s_2,...,s_{n-k})$ and $\vec{l}=(l_1,l_2,...,l_k)$ denote the stabilizer and logical quantum numbers respectively (both can be either $0$ or $1$). The state $\vec{r}=(r_1,r_2,...,r_k)$ are the standard $Z$ quantum numbers of the reference qubits. For instance, when $k=1$ the state shown in Eq.~\eqref{eq:initial_state} translates to $|0_L0_R\rangle \equiv |\vec{s}_0,0,0\rangle$ and $|1_L1_R\rangle \equiv |\vec{s}_0,1,1\rangle$. Where $\vec{s}_0=(0,0,...,0)$ is the configuration with all stabilizer eigenvalues are equal to $+1$. The stabilizer basis guarantees a \emph{more sparse} representation of the logical states than the initial $Z$ qubit basis. 

Any Pauli error $E \in \vec{P}$ ($\vec{P}$ is the Pauli group) with support only on the data qubits, acting on the basis elements of Eq.~\eqref{eq:basis} satisfies:
\begin{equation}\label{eq:basis_error}
E|\vec{s},\vec{l},\vec{r}\rangle = |\vec{s}',\vec{l}',\vec{r}\rangle.
\end{equation}
Pauli errors do not create \emph{coherent} superpositions in the stabilizer basis. This is a key insight since it follows immediately that CPTP maps $\mathcal{N}$ whose Kraus operators are elements of $\{\vec{P}\}$ do not create coherences in the stabilizer basis (we set $k=1$ from now on for the sake of simplicity):

\begin{equation}\label{eq:basis2}
\mathcal{N}\left(|\vec{s},\vec{l},\vec{r}\rangle \langle \vec{s},\vec{l},\vec{r} |\right) = \sum_{i,j} \gamma_{ij}|\vec{s}_i,\vec{l}_j,\vec{r}\rangle \langle \vec{s}_i,\vec{l}_j,\vec{r}|. 
\end{equation}
Here $\gamma_{i,j}$ are real numbers determined by the specific choice of $\mathcal{N}$ and the initial state $|\vec{s},\vec{l},\vec{r}\rangle$. This is the kind of CPTP maps that are used as error models. Now let us write down the mixed state $\rho_{RQ}$ in the stabilizer basis:
\begin{eqnarray}\label{eq:mixed_state}
\rho_{RQ}  & = & \mathcal{N}(|\Psi\rangle\langle \Psi|) \nonumber \\
           & = & \sum_{\vec{s}} \dfrac{\gamma^1_{s}}{2}  |\vec{s},0,0\rangle \langle \vec{s},0,0|+ \dfrac{\gamma^2_{s}}{2}  |\vec{s},1,1\rangle \langle \vec{s},1,1| \nonumber  \\
           && + \left(\dfrac{\gamma^3_{s}}{2}  |\vec{s},1,1\rangle \langle \vec{s},0,0|+h.c. \right) 
\end{eqnarray}
Since there are no cross terms $\vec{s}\vec{s'}$ in the summation, the density matrix $\rho_{RQ}$ is diagonal in the stabilizer quantum number $\vec{s}$. One can show that, in general, there are $2^{n-k}$ blocks of size $4^k$ each. Therefore the computation of entropies $S(\rho_{RQ})$ and $S(\rho_{Q})$ can be done block-wise in parallel. Thus the task is to store the mixed state $\rho_{RQ}$ in the stabilizer basis. In this work we do so by transforming the errors to the stabilizer basis and writing down the state directly in the same basis.

First we need to find an Clifford encoding unitary $U_{enc}$ such that $|0_L\rangle = U_{\text{enc}}|0\rangle^{\bigotimes n} $ and $|1_L\rangle = U_{\text{enc}}\overline{X}|0\rangle^{\bigotimes n} $. Here $\overline{X}$ is an operator that flips some qubits in such a way that the same encoding circuit can prepare both $|0_L\rangle$ and $|1_L\rangle$. This operator is always set by the specific choice of $U_{\text{enc}}$. In our case we find encoding circuits for each code using Stabgraph \cite{amaro_scalable_2020}.

The resulting state is then
\begin{eqnarray}\label{eq:mixed_state1}
\rho_{RQ}  & = & \mathcal{N}\left(U_{\text{enc}}|\psi\rangle\langle \psi|U^{\dagger}_{\text{enc}}\right) \\
           & = & \sum_{E} p(E) (E U_{\text{enc}}) |\psi\rangle\langle \psi| (U^{\dagger}_{\text{enc}} E^{\dagger})
\end{eqnarray}
Since $U_{\text{enc}}$ only has Clifford operations and $E$ are elements of the Pauli group, we can always do $E U_{\text{enc}} = U_{\text{enc}} \tilde{E} $. Then we re-expressed the state as:
\begin{eqnarray}\label{eq:mixed_state2}
\rho_{RQ}  & = & U_{\text{enc}} \left( \sum_{E} p(E) \tilde{E} |\psi\rangle\langle \psi|  \tilde{E}^{\dagger} \right) U^{\dagger}_{\text{enc}} \nonumber \\
          & = & U_{\text{enc}} \tilde{\mathcal{N}}(|\psi\rangle\langle\psi|)U^{\dagger}_{\text{enc}} \\
          & =  & U_{\text{enc}} \tilde{\rho}_{RQ}U^{\dagger}_{\text{enc}}
\end{eqnarray}
with $\tilde{\mathcal{N}}$ as the CPTP map with Kraus operators $\tilde{E}$ coming from the equation $E U_{\text{enc}} = U_{\text{enc}} \tilde{E} $. Since the coherent information is unaffected by coherent global rotations the following equality holds:
\begin{eqnarray}\label{eq:ci_pauli}
I(\rho_{RQ}) = I(\tilde{\rho}_{RQ}).
\end{eqnarray}
Effectively, $\tilde{\rho}_{RQ}$ is written in the stabilizer basis and $\tilde{E}$ are the errors transformed to that basis. Therefore it suffices to compute $\tilde{\rho}_{RQ}$ by sparse matrix-matrix multiplication and then read off the individual blocks. The coherent information is computed by adding up the contributions of each block. Importantly, this back-propagation of errors also applies to phenomenological and circuit level noise setups. The working basis is the stabilizer basis in the data qubits and the $Z$ basis in the ancilla and reference qubits. The error model consists of one and two qubit channels at specific locations, each error is then propagated back through all gates.

There is a further simplification for CSS codes and bit-flip noise model, namely that only $Z$ stabilizers are affected by the $X$ errors. This means that, out of $n-k=N_x+N_z$ ($N_{x(z)}$ is the number of $X(Z)$ stabilizers) stabilizers, only $N_z$ are prompt to errors. This can be exploited when writing down the density matrix in the stabilizer basis:
\begin{eqnarray}\label{eq:mixed_state}
\rho_{RQ}  & = & \sum_{\vec{s}_z} \dfrac{\gamma^1_{s}}{2}  |\vec{s}_z,0,0\rangle \langle \vec{s}_z,0,0|+ \dfrac{\gamma^2_{s}}{2}  |\vec{s}_z,1,1\rangle \langle \vec{s}_z,1,1| \nonumber  \\
           && + \left(\dfrac{\gamma^3_{s}}{2}  |\vec{s}_z,1,1\rangle \langle \vec{s}_z,0,0|+h.c. \right).
\end{eqnarray}
Now the summation runs only over the $Z$ stabilizers, the entries of the nontrivial $N_x$ quantum numbers are zero. In the case of surface and color code we have $N_x = N_z$ and $k=1$, therefore the total size of the relevant Hilbert space is $2^{(n-1)/2}$. This allows us to go to $d=7,9$ color code that requires $n=31,49$ data qubits respectively, and $d=7$ surface code that consists of $n=49$ data qubits. In order to do so one must rewrite the errors $\tilde{E}$ for acting only on the relevant Hilbert space. 

\subsection{Timing for computing the coherent information}

\begin{figure}[h]
    \centering
    \includegraphics[width=0.45\textwidth]{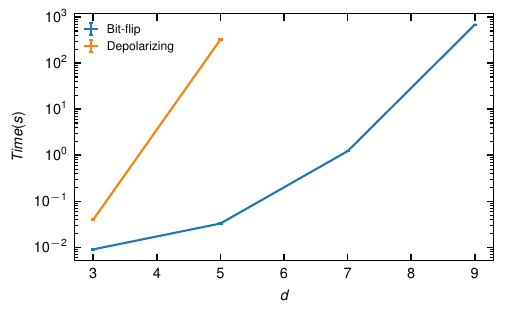}
    \caption{Time for computing the coherent information of the color code at fixed error rate for bit-flip and depolarizing noise. This is a single core computation on a Apple M1 Pro processor which has a maximum clock speed of $3$GHz.}
    \label{fig:timing}
\end{figure}

In this section we briefly discuss the computational time needed for computing the CI in the codes and error models used in this work. In Fig.~\ref{fig:timing} we show the time needed for computing the CI for the color code under bit-flip and depolarizing noise. All the simplifications exposed in the previous section are already implemented in the data shown. For the largest distance in both error models the time needed is approximately $\sim10$ minutes.

It is instructive to investigate the time needed with Monte Carlo integration of classical spin models for obtaining one data point $(O,p)$ where $O$ is, for example, magnetization. We can take the simulation parameters given in Table I of Ref.~\cite{katzgraber_error_2009}. The authors investigate the statistical mechanics mapping of the 6.6.6 color code under bit flip noise. Now we compare the time for both methods to obtain one data point $(O,p)$ where $O$ is the coherent information in the present work. After collecting the simulation parameters in \cite{katzgraber_error_2009} for $L=12$ and $p=0.04$, we estimate that $\sim 10^{10}$ Monte Carlo sweeps (including disorder samples and the different temperatures used for tempering) are needed for reaching equilibrium at the system sizes and temperature/error rate values presented in their work.
The runtime of a single Monte Carlo sweep depends on such a large amount of factors that is not possible to give a general estimation on. One could try to get a rough estimate by assuming a generous 15 FLOPS (50\% efficiency on the maximum clock speed which is quite over optimistic), a local Monte Carlo update needing 10 floating point operations (in reality could be orders of magnitude higher) and $L^2=144$ local updates to complete a sweep. Thus one point $(O,p)$ needs about $\sim 24 $ minutes for the smallest system size investigated. Of course, this number is far from rigorous and must be taken with a grain of salt, the actual number is most likely way higher. Thereby we see that both times are comparable even under over optimistic assumptions and larger system sizes for the Monte Carlo calculation. In conclusion, the CI method described in our work is at least as good as the statistical mechanics mapping in delivering optimal thresholds, with the key difference that the CI is, by definition, not tailored to a specific code and error model.

\end{document}